\documentclass[smallextended]{svjour3}
\usepackage{amsmath}
\usepackage{amsfonts}
\usepackage{amssymb}
\usepackage{graphicx}

\begin{document}

\title{Failure and Uses of Jaynes' Principle of Transformation Groups}

\author{Alon Drory}
\institute{Afeka College of Engineering, 218 Bney-Efraim Street, Tel-Aviv, 69107 , Israel}
\email{adrory@gmail.com}

\begin{abstract}
Bertand's paradox is a fundamental problem in probability that casts doubt on the applicability of the indifference principle by showing that it may yield contradictory results, depending on the meaning assigned to ``randomness''. Jaynes claimed that symmetry requirements (the principle of transformation groups) solve the paradox by selecting a unique solution to the problem. I show that this is not the case and that every variant obtained from the principle of indifference can also be obtained from Jaynes' principle of transformation groups. This is because the same symmetries can be mathematically implemented in different ways, depending on the procedure of random selection that one uses. I describe a simple experiment that supports a result from symmetry arguments, but the solution is different from Jaynes'. Jaynes' method is thus best seen as a tool to obtain probability distributions when the principle of indifference is inconvenient, but it cannot resolve ambiguities inherent in the use of that principle and still depends on explicitly defining the selection procedure.

\end{abstract}

\maketitle

\section{Introduction}
\label{sec:intro}

The application of probabilities in physics raises a host of difficult foundational questions \cite{sklar}. Among these is the problem of determining the correct ``counting'' procedure for a given case, or equivalently, choosing which probability distribution to use for the relevant random variables. Perhaps the most common approach is to use the principle of indifference (so named by Keynes, and often known previously as the principle of insufficient cause):

\begin{quotation}
\textbf{Principle of Indifference}: If there is no \textit{known} reason for predicating of our subject one rather than another of several alternatives, then relatively to such knowledge the assertions of each of these alternatives have an equal probability.  \cite[p.42]{keynes}
\end{quotation}

The principle of indifference, in some form or other, has been successfully used in myriad applications, from coin flipping and gambling games to configurations counting in statistical mechanics.  Nevertheless, its philosophical status and proper use are still widely debated. One of these debates originates in the work of mathematician Joseph Louis Fran\c{c}ois Bertrand (1822-1900), who created a series of examples showing how the principle can lead to trouble if applied uncritically. Later dubbed `paradoxes', the most famous of these is the chord problem, which has been discussed ever since (\cite{northrop,garwood,tissier,fraassen,gillies}).

\textbf{Bertrand's Problem}: Consider a circle with an equilateral triangle inscribed in it. What is the probability that a chord selected at random will be longer than the side of the triangle? 

Bertrand showed that one obtains (at least) three different answers according to the procedure used to select the chord.

\textbf{B1.}   Draw the triangle from the point marking one of the ends of the chord. The chord will be longer than the triangle's side if it is inscribed inside the angle at the vertex. This represents a third of the possible range for the angles between the chord and the tangent at the vertex, $\left[ 0 , \pi \right]$. The probability is 1/3, therefore.

\textbf{B2.}  A chord is completely defined by its midpoint. If the midpoint lies closer to the center than half a radius, the chord will be longer than the side of the triangle. This yields a probability 1/2.

\textbf{B3.}   All the midpoints of chords longer than the side of the triangle must fall within a circular area around the center of the original circle, of width half the original diameter. This area covers a quarter of the original circle and if the midpoints are chosen at random in the area of the circle, the probability of them falling in the relevant region is thus 1/4.

In each of these cases, we apply the principle of indifference, but obtain a different prediction.

Approaches to Bertrand's paradox vary. Some authors seek to deny that the problem is truly a problem [for a recent example, see \cite{bangu}; for a criticism of this position see \cite{row}]. Probably the most common attitude is to show that the problem is not well-posed, i.e., that it conflates several different problems because ``random" is an imprecise term. Once the various problems are disentangled and properly understood, the principle of indifference yields a unique prescription for the probability. The classic presentation of this approach is given by Marinoff \cite{marinoff}. Although this was apparently Bertrand's own position, there has been some debate whether this represents an actual resolution of the paradox \cite{shackel}.

In contrast, another approach claims that Bertrand's problem is actually well-posed and that it does have a unique solution, the multiplicity of answers being only apparent. The argument originates with Poincar\'{e} \cite{poincare} but the main proponent of this position is considered to be Jaynes \cite{jaynes}. His claim is that one should extend ``indifference'' in the principle to every aspect that is left unstipulated in the problem. Thus, the orientation, size and position of the circle should not matter. Jaynes takes this to mean that the required probability distribution function (PDF) of the chords should be invariant to changes in these properties, i.e., invariant under a group of transformations that includes rotations, rescaling and translations of the circle. This is his principle of transformations group.

Jaynes then performed an experiment in which he threw long straws at a circle drawn on the floor. The distribution of the chords generated by the intersections of the straws with the circle was then compared to the theoretical prediction and found to be consistent with it. This combination of experimental verification and theoretical justification has held great appeal ever since, particularly among physicists. The notion that there is a unique solution to Bertrand's problem that can be discovered experimentally has been revisited and extended several times since \cite{holbrook,diporto1,diporto2,wang1,wang2}.

It is widely accepted that Jaynes proved that there is a unique solution that possesses rotational, scaling and translational invariance \cite{fraassen,wang2,weisstein}. Even authors who reject Jaynes' approach nevertheless seem to agree that the symmetry requirement does lead to a unique solution, although they may dispute the validity of the requirement itself \cite{marinoff,gillies}. 

My aim here is to show that is not the case and that as with the principle of indifference, the application of the principle of transformation groups depends upon the method of selection of chords. The implementation of the symmetries turns out not to be unique, and leads to different mathematical requirements depending on the underlying process by which one imagines the random chords to be generated. In fact, contrary to Jaynes' assertion, each of the classical three solutions of Bertrand's problem (and additional ones as well!) can be derived by the principle of transformation groups, using the exact same symmetries, namely rotational, scaling and translational invariance.

I begin by slightly rephrasing Bertrand's problem in section \ref{sec:regular} in order to eliminate some criticisms of the standard formulation. This regularized version is used throughout the rest of the paper. I review Jaynes' application of his principle of transformation groups in section \ref{sec:jaynes}. In sections \ref{sec:probquart} and \ref{sec:probthird}, I then show that Jaynes' principle can also yield Bertrand's two other solutions. Section \ref{sec:release} presents a variation on the first solution and describes an empirical verification of the result. Section \ref{sec:conclusions} then discusses the implications of these facts for physical applications of the principle of indifference and what Jaynes' principle actually contributes to the problem.

\section{Regularized Bertrand's Problem}
\label{sec:regular}

Bertrand's original formulation has been the subject of several criticisms, and the validity of his solutions has been debated, on the grounds that the suggested procedures either do not select a chord from the set of all possible chords \cite{rowbottom}, or else that they fail to select a unique chord \cite{shackel}. Shackel, in particular, eliminates Bertrand's proposed solution B3, randomly selecting the midpoint in the area of the circle, because if the midpoint happens to be the center of the circle, it does not define a unique chord. Any diameter, no matter its orientation, is a possible chord. Shackel argues that this disqualifies the procedure\footnote{One must be historically fair to Bertrand, however. Although it has become standard to seek the probability that a chord is longer than the side of the inscribed triangle, Bertrand's own version required the probability that it be shorter. This obviously eliminates diameters from the sample space. Furthermore, it also proves that the standard solutions are all valid, since one can calculate the probability that a chord is longer than the side as one minus the probability that is it shorter. When applied to calculating the probability that a chord is shorter than the side, the argument underlying B3 is obviously valid and gives 3/4, which in turns yields indeed 1/4 for the probability that the chord is longer. This shows that the solution is actually sound. Nevertheless, I adopt a different approach here in order to leave no doubt about the matter.}.

Rowbottom, on the other hand, criticizes Bertrand's solutions as selecting chords from proper subsets of all chords instead of the whole set of possibilities \cite{rowbottom}. In his view, this disqualifies all the purported solutions. He admits that one can restate the problem to eliminate this issue, but argues that the rephrased problem is much harder and that Bertrand's solutions are not adequate answers.

The historical intentions of Bertrand are of no concern here, and the criticisms can be answered by slightly reformulating the question and the solutions. The essentials remain unchanged in this regularized version of Bertrand's paradox, Jaynes' solution is practically identical and all the questions raised are still present. Consequently, I will refer in this paper to the regularized problem and its solutions as Bertrand's paradox, without further specifications.

To get rid of the singularity at the center of the circle, let us rephrase the question as:

\textbf{Regularized Bertrand's Problem}: In a circle, select at random a chord \textit{that is not a diameter}. What is the probability that its length is greater than the side of the equilateral triangle inscribed in the circle?

The regularized versions of Bertrand's solutions are now as follows:

\textbf{RB1.} Select at random a point on the circumference of the circle. From that point, select at random an angle for the direction of the chord, excluding the direction of the diameter. If we use the diameter passing through the selected point on the circumference as the polar axis, we are selecting a random angle from the set $\left( -\pi/2 , 0 \right) \cup \left( 0,  \pi/2 \right)$, or equivalently from the range $\left( -\pi/2, \pi/2 \right) - \left\{0 \right\}$, i.e., excluding the angle $0$ corresponding to the diameter. For a given point on the circumference, the range of angles yielding a chord longer than the side of the triangle is $\left(- \pi/6, 0 \right) \cup \left(0, \pi/6 \right)$, yielding a probability of $1/3$. Since all the points on the circumference are equivalent, the probability of any chord which is not a diameter being longer than the side of the triangle is also $1/3$ (see also section \ref{sec:probthird} for a more formal discussion).

\textbf{RB2.} Randomly select a diameter. Consider the radius that is perpendicular to the selected diameter. Select at random a point on this radius and draw the chord that passes through it, which is also parallel to the selected diameter. If the distance of the point from the center lies in the open range $\left(0 , R/2 \right)$, the corresponding chord will be longer than the side of the triangle. Thus, for any given diameter, the probability of selecting such a chord is $1/2$. Since all the firstly selected diameters are equivalent, the probability of any chord (which is necessarily parallel to \textit{some} diameter) being longer than the side of the triangle is also $1/2$.

\textbf{RB3.} Select at random a point inside the circle, excluding its center. Any such point selects a unique chord of which it is the midpoint. The chord will be longer than the side of the triangle if the selected point falls inside a central circle of radius $R/2$, excluding the center itself. The surface of this area is still $\pi R^2/4$, since the excluded central point has surface zero, and the corresponding probability for the appropriate chord is thus $1/4$.

All these methods select one single chord, from the set of all possible chords that are not diameters, and they all require exactly two random variables to fully specify the chord. Thus they all represent (at least a priori) equally acceptable applications of the principle of indifference to the solution of the regularized Bertrand's problem.

With respect to this regularized problem and solutions let us now consider Jaynes' arguments for the unique correctness of the value $1/2$ as \textit{the} solution of the problem. As mentioned above, for the sake of conciseness I will continue to speak of Bertrand's problem and solutions when I mean the regularized versions of these. For the same reason I will not mention every time that diameters are excluded from our selection, unless it has direct implications for the calculation at hand. I shall specify how Jaynes' analysis has to be adapted to the regularized problem, however. As we shall see, it requires practically no change.

\section{Jaynes' Solution: Probability 1/2}
\label{sec:jaynes}

Jaynes seeks the probability distribution of chords that satisfies certain symmetries of the problem. To do so, he needs to select mathematical parameters that characterize completely a chord. As we shall see later, this proves to be a crucial point. Jaynes chooses the polar coordinates of the chord's midpoint, $\left( r, \theta \right)$, relative to an origin located at the center of the circle. $f(r,\theta) dS$ is the probability of having the midpoint in the infinitesimal area $dS = r dr d\theta$.

Jaynes then applies his principle of transformation groups. In this instance, it means that the required PDF must be identical for all observers that are not explicitly distinguished in the formulation of the problem. Bertrand's problem contains no restrictions on the orientation, location or scale used by the observers, and therefore the required PDF should be symmetrical with respect to rotations, scale changes and translations. Note that the transformations are supposed to be applied to the observers (or rather to their coordinate systems), not to the chords themselves.

\subsection{Rotational Symmetry}

Consider two observers and let the polar axis of observer B be rotated clockwise by an angle $\alpha$ with respect to observer A's axis. Observer A describes the system with a PDF $f\left( r, \theta \right)$, while observer B assigns to the problem a (potentially) different PDF $g\left( r , \theta \right)$. Now, an angle assigned the value $\theta$ by observer A corresponds to an angle $\theta - \alpha$ in the system of B. The system itself is identical, only its descriptions by the two observers differ. Hence, the two PDF's must be identical when referring to the same situation, which means that
\begin{equation}
\label{eq:rot}
f\left(r, \theta \right) = g\left( r, \theta - \alpha \right)
\end{equation}

This equation merely expresses the arbitrariness of the axes used to describe the situation, whether the problem itself exhibit rotational invariance or not. 

Jaynes then argues that the Bertrand problem contains no restriction on the orientation of the chords, which implies that the solution must be identical for observers that are merely rotated with respect to each other. This invariance implies that $f\left(r, \theta \right) = g\left( r , \theta \right)$, which in turn, implies, because of Eq.(\ref{eq:rot}), that $f\left(r, \theta \right)$ must be independent of $\theta$. In other words, $f\left(r, \theta \right) = f(r)$.

\subsection{Scale Invariance}

Consider two observers using different scales so that observer A measures a circle of radius $R$, while observer B measures a circle of smaller radius $aR$, with $a \leq 1$ being the scale factor. The two circles are concentric, scale being the only difference. Again, let observer A describe the system with a PDF $f(r)$, while observer B uses a PDF $h\left(r \right)$. If the midpoint of the chord falls inside the small circle, it simultaneously defines a chord on the smaller circle and on the larger circle. This means that for $r \leq a R$, the two distributions $f(r)$ and $h\left(r \right)$ must be proportional to each other. They are not equal because they differ in their normalization, since $f(r)$ also accounts for chords where $r > aR$, which are excluded from $h\left(r \right)$. Thus, $h\left(r \right)$ can be thought of as conditional probability, i.e.,
\begin{equation}
\label{eq:conditional}
h(r) = \dfrac{f(r)}{Prob(\text{midpoint falls inside smaller circle})}
\end{equation}
which can now be rewritten (taking into account rotational invariance) as:
\begin{equation}
\label{eq:h(r)}
h(r) = \dfrac{f(r)}{\int_0^{aR} f(r) 2\pi r dr}
\end{equation}
This relation holds whether the system is scale invariant or not. It merely represents the transformation of one probability density function into the other when we change scale. Note that in this and every subsequent integral over distances, the single point $r=0$ is excluded from the range of the integral. But since all the integrands involved are regular, this single point makes no difference to the value of the integral itself. Thus, Jaynes' argument and its result apply equally to the regularized Bertrand's problem as well as to the original version. 

Jaynes now invokes again epistemic indifference. Since the size of the circle is left indeterminate in the problem, the solution ought to be independent of it. This means that if an observer rescales all his distances by a factor $a$ (i.e., $r \longrightarrow a r$), the resulting distribution ought to remain as before, so that 
\begin{equation}
\label{eq:scaleinv}
h(ar) (ar) d(ar) d\theta = f(r) r dr d\theta   \longrightarrow  a^2h(ar) = f(r)
\end{equation}
Replacing $r$ by $ar$ in Eq.(\ref{eq:h(r)}) and substituting the relation Eq.(\ref{eq:scaleinv}) yields the integral equation
\begin{equation}
\label{eq:integscale}
a^2f(ar) = 2 \pi f(r) \int_0^{aR} f(u) u du
\end{equation}
where again I will mention for the last time that the point $u=0$ is excluded from the range.

By differentiating with respect to a, one can transform this into a differential equation. Its solution is:
\begin{equation}
\label{fr}
f(r) = \dfrac{q r^{q-2}}{2 \pi R^q}
\end{equation}
where q is an arbitrary constant.
	This distribution is consistent with two of the three Bertrand solutions. Solution RB2 (as well as B2) corresponds to q = 1 and solution RB3 to q = 2. Solution RB1 (or B1) is therefore eliminated at this stage.

\subsection{Translational Invariance}

To determine the value of the parameter q, Jaynes invokes translational invariance. Since the position of the circle's center is not one of the random variables in the PDF, imposing this symmetry requires some interpretation and this is precisely where Jaynes' analysis must rely on a specific method of chord selection. Jaynes always thinks of chord selection through his experimental implementation of throwing long straws at a circle inscribed on the ground. Translational invariance then takes a very specific meaning. The straw's throw is unrelated to the position of the circle. Provided the straw is infinitely long, it defines a line along which the chord lies. The chord and its midpoint $\left( r , \theta \right)$ are determined through the intersection with the circle, and if the circle is translated, the chord and its midpoint are also translated to new values $\left( r' , \theta' \right)$. The situation is illustrated in Fig. 1, where a single line intersects two circles, C and C', the second being displaced by a distance $b$ with respect to the fist. That intersection generates two different chords, one for each circle, and the midpoints of these chords are \textit{different} points. Let $r$ and $r'$ be the distances of the chords' midpoints to their respective circle centers and $\theta$ and $\theta'$ the directions of these midpoints. Then we have that 
\begin{eqnarray}
\label{rprime}
r' = \lvert r - b cos \theta \rvert \nonumber\\
\theta' = \left\{ \begin{array}{ll}
	\theta & r > b cos\theta \\
	\theta + \pi & r < b cos\theta
	\end{array} \right.
\end{eqnarray}

\begin{figure}
\label{fig:jaynestrans}
\includegraphics[scale=0.13]{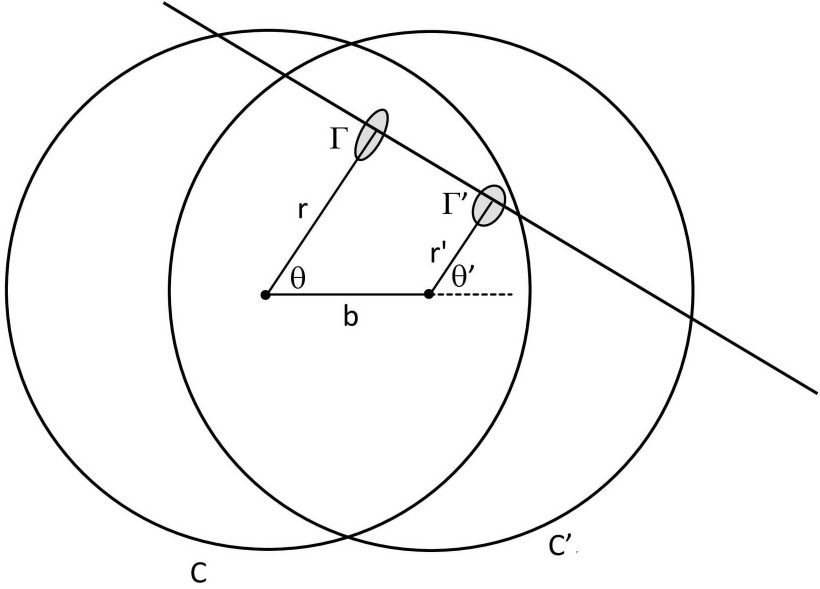}
\caption{Translational invariance in Jaynes' straw throwing procedure. A single straw may intersect two circles, translated by a distance $b$, thus generating two chords. The midpoints of the chords are then also translated with respect to each other and lie at different distances from the centers of their respective circles}
\end{figure}

As we throw more straws, the midpoints of the chords generated in the original circle may vary over an area $\Gamma$. These same lines then generate chords in the translated circles, whose midpoints vary over a different area, $\Gamma'$, which need not be identical to $\Gamma$. In fact, the infinitesimal surface element $dS = r dr d\theta$ is itself different from the surface element $dS' = r'dr'd\theta$.

Jaynes then argues that translational symmetry implies that the probability of the midpoints of chords in the original circle lying in the area $\Gamma$ must be identical to that of the midpoints of chords in the translated circle, which lie in the area $\Gamma'$. In other words, the mathematical implementation of translation invariance is
\begin{equation}
\int_{\Gamma}f(r) r dr d\theta = \int_{\Gamma'}f(r') r' dr' d\theta'
\end{equation}

Substituting now the result Eq.(\ref{fr}) for the form of $f(r)$, and using the definitions of $r'$ and $\theta'$, Eq.(\ref{rprime}), to transform the right hand side term, we obtain
\begin{equation}
\int_{\Gamma}r^{q - 1} dr d\theta = \int_{\Gamma}\lvert r - b cos\theta \rvert^{q - 1} dr d\theta
\end{equation}
where we used the fact that the jacobian of the transformation from $(r, \theta)$ to $(r', \theta')$ is unity. 

This relation implies that $q = 1$, and the resulting PDF $f(r)$ yields the value $0.5$ for the probability of a chord being longer than the side of the inscribed triangle. This value is consistent with the solution RB2 (and B2), though it is unclear at first sight that it does indeed represent an implementation of this solution. RB2 selects the midpoint of the chord directly, but although Jaynes' procedure uses the midpoint as a \textit{characterization} of the chord, it does not seem to contain a physical \textit{selection} of the midpoint itself. It might be the case, therefore, that it represents a different procedure from all of Bertrand's classical solutions, which only coincidentally yields the same numerical value as RB2. Indeed, Chiu and Larson found that several different methods of chord selection may yield the same Bertrand probability \cite{chiu}. In this case, however, even though Jaynes' procedure is not a direct implementation of RB2, there is reason to believe that the numerical agreement is not coincidental. Tissier gives a geometric argument that suggests why the two procedures agree (although he does not prove the agreement formally) \cite{tissier}. The argument is based on the fact that since the PDF $f(r)$ only depends on the distance $r$, one can rotate all the lines as long as $r$ remains constant, and redraw them all as parallel lines. This maps Jaynes' procedure onto a different procedure in which all the chords are selected from a set of parallel lines, which is precisely the procedure underlying RB2. 

It would seem now that we have proved that Bertrand's problem was well-posed after all. Using the symmetries implicit in the formulation of the problem, we have obtained a single solution. Moreover, this formulation fits an experimental method of selection of chords and the results of Jaynes' analysis are thus empirically confirmed. What more could we want? 

But precisely the experimental ``confirmation'' of the analysis proves to be its undoing. We have the logical order of things backwards. The straw throwing experiment is not the empirical confirmation of an independent abstract analysis. Instead, the analysis is a mathematization of the straw throwing procedure. This means that the experimental procedure logically precedes the analysis. It is on the basis of this procedure that the mathematical relations expressing the symmetries are derived. Far from being a confirmation that the principle of transformation groups yields indeed a unique solution, it is the choice of a specific experimental procedure that determines how the symmetries of the problem are mathematically implemented. As we shall now see, if we had thought of a different experimental procedure in the first place, the principle of transformation groups would have yielded a different result. The principle does not determine that a unique method of chord selection is correct, therefore, but rather the choice of chord selection procedure determines how the principle is applied. I shall now show how, contrary to Jaynes's opinion, the principle of transformation groups does yield Bertrand's other solutions.

\section{Throwing darts: Probability 1/4}
\label{sec:probquart}

The implementation of translational invariance in Jaynes's solution differs fundamentally from that of the previous symmetries. In both rotational and scale invariance, the center of the chord remains unchanged and is represented by the same point for the original circle as for the rotated or scaled circle. This is consistent with Jaynes' initial definitions. Note that Jaynes begins his analysis by seeking the distribution function for the midpoints of chords, defined by their position $(r, \theta)$. That distribution function, $f(r, \theta)$, must satisfy some mathematical constraints because different \textit{observers} attribute different values of $r$ and $\theta$  to the \textit{same} point, according to the axes and units that they use. Indeed, in sections 1 and 2 of his paper, Jaynes specifically refers to differences in \textit{observers} as underlying the invariance properties. Thus, two observers may use different axes to represent a single situation, and their descriptions must be consistent with each other. 

But the translational invariance imposed by Jaynes is of a completely different nature. The transformation equation does not arise from a difference in the observers, but rather from a requirement of invariance under the \textit{physical} translation of the circle. 

Now such ambiguity was already present in the scale invariance argument. There as well, Jaynes considered two different circles, rather than two observers. But one could claim that it made no difference. To any change of units by the observer there corresponds a physical change of scale of the system that yields the same situation. That argument depends, however, on the fact that both observers agree on the point that represents the chord's midpoint. 

The translation argument is different, because after the translation of the circle, the chord's midpoint is a physically different point. $r$ and $r'$ are not two different characterizations of the chord's midpoint; they represent the distances to two altogether different points (the two circles' centers). If Jaynes' procedure were a naively faithful representation of his choice of random variables, we should expect the chord's midpoint to be directly selected, so that whatever point we chose \textit{must} be the chord's midpoint by definition. That is not the case. The circle's translation results in a mapping of the chord's midpoints from one area to another. As a result, the areas $\Gamma$ and $\Gamma '$ are different. This is the reason that Jayne's condition implies that $r f(r, \theta)$ is a constant function, rather than, for example, $f(r, \theta)$ itself.

There is of course nothing wrong with stipulating, as Jaynes does, a specific process by which the coordinates of the chord's midpoint are selected even if the selection is only indirect. And it is perfectly possible and valid that a change in the conditions of the selection, such as a translation of the circle, should result in a different midpoint being selected. But neither is it natural or inevitable. It is no more than a particular characteristic of a specific process.

Indeed, one can implement the very same symmetry, translational invariance, in a different way, if one assumes the midpoint selection to be determined by another process. Specifically, imagine that we wish to directly select a point that is \textit{defined} to be the chord's midpoint. One such selection process (perhaps not the most ``natural'' one) could be for example to have straws impaled on darts, so that the dart serves as an axis around which the straws can rotate freely. We then throw such darts at a circle. Provided the dart hits the inside of the circle (excluding its center), it is defined to be the midpoint of a chord. The straw is then rotated around the dart until the chord it generates is centered on the dart. There is only one such chord, so the selection is well-defined. In this procedure, unlike Jaynes', the midpoint is selected first and the chord is determined by it, rather than the other way around. When one throws straws, in other words, one selects a chord and the midpoint is determined from the chord. When one throws darts, one selects a midpoint and the chord is determined from that. Note that in \textit{neither} this nor Jaynes' procedure can we directly select both the chords and the coordinates of their midpoints.

Now consider the situation depicted in Fig. 2, in which we compare again two circles, $C$ and $C'$, whose centers are translated by a distance $b$. Assume the dart hits the point $O$, a distance $r$ from the center of the circle $C$ and a distance $r'$ from the center of the circle $C'$. In the circle $C$, this generates the chord $AB$, centered around $O$. In the circle $C'$, however, $O$ is at a different distance from the center, and the chord whose midpoint is $O$ is $A'B'$, which differs from $AB$. This is the exact counterpart of Jaynes' procedure, in which one single line generates the chord in both circles and the midpoints are different. Here, the midpoint remains constant and the chords are different.

\begin{figure}
\label{fig:darttrans}
\includegraphics[scale=0.14]{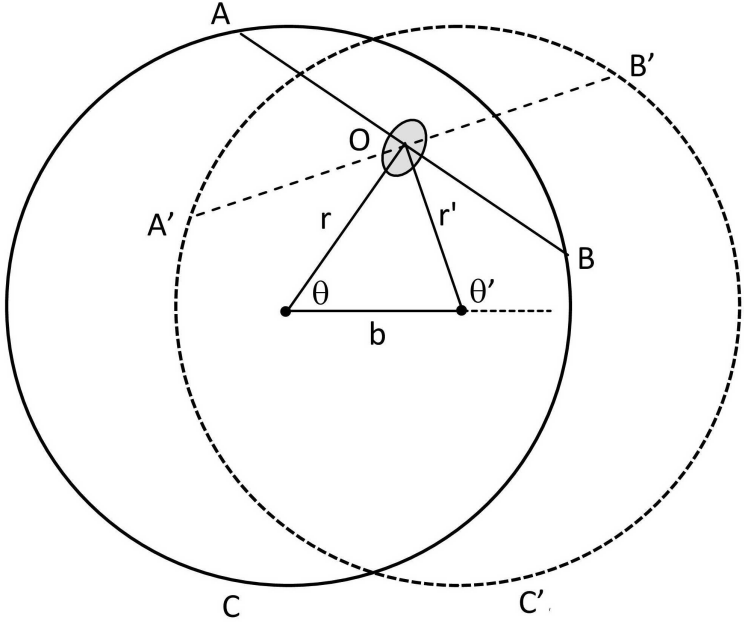}
\caption{Translational invariance in the impaled straw procedure. The selected point (where the dart lands) is defined to be the midpoint of the chord. If two circles are translated by a distance $b$, the same midpoint will belong to two different chords.}
\end{figure}

Just as before, invariance under translation requires that
\begin{equation}
\int_{\Gamma}f(r, \theta) dS = \int_{\Gamma'}f(r', \theta') dS'
\end{equation}

The difference between this case and Jaynes' is that under the dart-throwing procedure, the areas $\Gamma$ and $\Gamma'$ are identical, because we select the midpoints directly. As a matter of fact, $dS$ and $dS'$ themselves are identical, since the points that occupy them are identical. As a result, the condition of translational invariance now requires that 
\begin{equation}
\label{eq:transalt}
f(r, \theta) = f(r', \theta') \Longrightarrow  r^{q - 2} = \left( r' \right)^{q - 2}
\end{equation}
where $r'$ is given by the relation
\begin{equation}
r' = \sqrt{r^2 + b^2 - 2 r b cos(\theta)}
\end{equation}
Eq.(\ref{eq:transalt}) then implies immediately that $q = 2$ and that $f(r, \theta) = \dfrac{1}{\pi R^2}$, a constant. The probability of a chord's midpoint lying in an area $S$ is then proportional to that area, and the Bertrand probability is $\dfrac{1}{4}$, confirming RB3.

This result was obtained through Jaynes' principle of transformations group, just like Jaynes' original result was, yet the two results are different. Obviously, and contrary to Jaynes' claim, the principle does not select a unique solution for the probability distribution function. The reason is that the mathematical implementation of the symmetries is not unique. Translation, for example, is not a uniquely defined set of transformations under which the probabilities must be invariant. As it turns out, there are several possible mathematical expressions of the \textit{fact} of invariance under translation, and these depend on the experimental procedure selected to generate the PDF. Thus, Jaynes' principle will yield either Bertrand's solution RB2 or the solution RB3, depending on which equations we set up to express the fact of translational invariance. Throwing straws to select chords suggests a particular mathematical implementation of translational invariance. Throwing darts to select chords' midpoints suggests a different one. Neither is more intrinsically correct than the other. Each represents the conditions of the chosen experimental procedure.

In fact, however, the multiplicity of solutions is even wider, for solution RG1 is also derivable from Jaynes' principle, using precisely the same symmetries – namely, rotation, scale and translation invariance. This seems odd at first, because Jaynes seemingly proved that this solution violates scale invariance. This turns out to be misleading, however. The reason is that right at the start, Jaynes arbitrarily decided to describe the chord through the coordinates of its midpoint. But this characterization is not unique and one could equally well describe a chord using a different set of variables. Applying the principle of transformation groups to those variables generates a different set of mathematical conditions, which in turn yield a different solution for the PDF.

\section{Following spinners: Probability 1/3}
\label{sec:probthird}

Jaynes' application of transformation groups rejected Bertrand's solution RG1 because it violates scale invariance. Yet the textual description of procedure RG1 includes no obvious scale-dependent element, and it thus seems mysterious that it should be eliminated on these grounds. As I shall show now, this is nothing more than an artefact of Jaynes' own choice of the procedure of chord generation, and a different choice can make this solution viable.
 
The regularized procedure RB1 contains two steps and can be implemented empirically by way of a spinner, such as a needle mounted on a vertical axis so that it is free to rotate in the horizontal plane. The experimenter starts at the center of the circle and spins the needle. He then proceeds in the direction in which the needle points after coming to rest, until he intersects the circle. This selects a random point on the circumference. From that point, the experimenter spins his needle again and draws a line along the direction it indicates. Now in half the cases, the needle will point away from the circle. To overcome this problem, let us decree that the experimenter draws the line in both directions from his position, so that one such extension is certain to fall inside the circle and define a chord.  

Formally, we identify the two directions determined by the spinner by two angles $\alpha$ (the direction of the radius-vector to the chosen end-point on the perimeter) and $\beta$ (the direction of the chord from that point). We also specify a sign convention for both angles, such as their being positive if measured counterclockwise. We are now seeking the PDF of these angles, $f_1\left( \alpha , \beta \right)$, as determined by an appropriate group of transformations (always excluding diameters).

One should note that our procedure does not map a chord to a single pair of angles, but rather to four such pairs. Indeed, since a chord has two end-points, there are two possible values of $\alpha$ for every single chord (denoted, e.g., $\alpha_1$ and $\alpha_2$), each corresponding to one of the end-points. Moreover, our procedure dictates that the angles $\beta$ and $\beta + \pi$ correspond to the same chord, one of these angles pointing into the circle and the other pointing outwards. Since we decided to extend our line in both directions, both these angle generate the same line in the plane, and thus the same chord. 

The mathematical relations between the four sets of angles corresponding to a single chord depend on the direction with respect to which $\alpha$ and $\beta$ are measured. For example, if $\beta$ is measured as the angle between the chord and the radius-vector to the endpoint, then the chord determined by $\left( \alpha , \beta \right)$ is identical to that determined by the pairs $\left( \alpha , \beta + \pi \right)$, $\left(\pi +\alpha - 2 \beta, - \beta \right)$ and $\left( \pi + \alpha - 2 \beta , \pi - \beta \right)$. Other definitions generate different expression for the four sets of corresponding angles. The freedom to determine the directions with respect to which $\alpha$ and $\beta$ are determined turns out to be crucial to the problem at hand, as we shall shortly see.

Although this 1:4 correspondence might be a problem in some formalizations, it is actually of no import as far as probabilities are concerned. This is because $f_1 \left( \alpha , \beta \right)$ counts \textit{every} chord \textit{exactly} four times\footnote{This is because we exclude the diameters, which correspond to $\beta = 0$, so that $\beta$ and $- \beta$ always correspond to different angles.}, and it is normalized with respect to the total number of choices corresponding to all possible values of $\alpha$ and $\beta$, which is also exactly four times the total number of chords in the circle, excluding diameters\footnote{I use the term ``number'' informally here, for brevity's sake. One can easily transfer this to the language of measures.}. Therefore, the factor 4 cancels out and the probabilities are identical whether we count every chord once or four times. 

The change in procedure implies a change in the application of the relevant transformation groups. Since the PDF $f_1 \left( \alpha , \beta \right)$ contains only angular variables, it is automatically scale invariant, so this symmetry adds no new information (note that this was the symmetry that eliminated this solution in Jaynes' original analysis). 

Translations are also of little help because the angles are only meaningful with respect to a determined, though arbitrary, center. Another way of saying this is to consider the pragmatics of the experimental procedure. In Jaynes' method, one throw straws onto the ground and this operation can be performed without reference to the circle (provided that the straws are effectively infinite). In other words, the experimenter need not know where the circle is located in order to perform the procedure. Not so in the present case. The first spin determines the direction in which the experimenter moves from the circle's center. Without knowing the position of this center, no chord can be selected. This does not contradict Jaynes' argument that the position of the circle should not matter. It does not matter here as well. The circle's center can be translated at will. But the procedure only starts \textit{after} that position has been fixed. Thus, although the circle's location is arbitrary (as required in Jaynes' analysis), it must be determined before the chord is selected. 

Translations have no import for the form of the PDF we seek, therefore, because this form is not altered if the circle is translated \textit{previous} to the selection of $\alpha$ and $\beta$, and the procedure itself forbids any such translation \textit{after} the selection of the angles. Thus, although the scale and location of the circle are as arbitrary here as they were in Jaynes' procedure, they have no influence on the form of the PDF and thus yield no information with regards to it.

Nevertheless, there is one element that remains undetermined in the procedure, and this single symmetry fixes the PDF completely. Indeed, at no point did we specify which way the experimenter was facing while spinning the needle and the result should therefore be independent of this information. This is a type of rotational symmetry, which can be formalized in the following way. The direction in which the experimenter is facing can be taken as the zero of the needle's angle, i.e., the direction of the polar axis with respect to which it is measured. To say that Bertrand's problem leaves this direction unspecified is equivalent to saying that we are free to choose the direction of the polar axis arbitrarily. But in our procedure the experimenter spins the needle twice and there are no reasons to connect his orientation the first time with that which he adopts the second time. In other words, there is no necessity to measure both angles with respect to the same axis. As long as the two axes are known, thus unequivocally determining the spatial directions, they can be chosen independently of each other. This is because we never compare $\alpha$ and $\beta$ directly or combine them during the chord selection\footnote{The only exception to this is the relation between the four angles that we identify as selecting the same chord. These are dependent on the axes chosen. However, this does not influence the independence of the choice of angles and axes in the first place.}. Indeed, as noted above, $\beta$ could be defined as the angle between the chord and the radius-vector, whereas $\alpha$ must obviously be defined with respect to some other axis. An observer B can use axes rotated by angles $\theta$ (for the axis defining $\alpha$) and $\phi$ (for the axis defining $\beta$). The angles he uses are then
\begin{eqnarray}
\alpha' = \alpha - \theta \nonumber \\
\beta' = \beta - \phi
\end{eqnarray}

The empirical situation is independent of the choice of axes, and the problem contains no restriction on this choice. Using Jaynes' own argument, this lack of restrictions implies that the solution must be identical for both observers. We must have that
\begin{equation}
f_1\left( \alpha, \beta \right) d\alpha d\beta = f_1(\alpha', \beta')d\alpha'd\beta'
\end{equation}
or in other words,
\begin{equation}
f_{1}(\alpha,\beta) = f_{1}(\alpha - \theta, \beta - \phi)
\end{equation}
Since $\theta$ and $\phi$ are arbitrary, and considering the normalization condition, the only solution to this relation is the constant function:
\begin{equation}
f_1 \left( \alpha, \beta \right) = \dfrac{1}{4\pi^2}
\end{equation}
The corresponding answer to Bertrand's problem is now:
\begin{equation}
\label{eq:spinprob}
P_{1} = \int_{0}^{2 \pi} \int_{-\pi/6}^{\pi/6}f_1(\alpha, \beta) d\alpha d\beta + \int_{0}^{2 \pi} \int_{5\pi/6}^{7\pi/6}f_1(\alpha, \beta) d\alpha d\beta = \dfrac{1}{3}
\end{equation}
where we defined the angle $\beta$ as the angle between the chord and the radius vector. Thus we have obtained solution RB1 by using the exact same symmetries Jaynes used in his original analysis to justify RB2. This is because the different selection procedure we defined yields different mathematical expressions of the symmetries. Thus we see again that the principle of transformation groups does not cut through the multiplicity of solutions suggested by the principle of indifference. Instead, it has just as many various interpretations itself, dependent on the chord generation procedure, because a ``symmetry'' does not correspond to a unique mathematical prescription but rather to a general recipe for generating such a prescription, based on the manner in which the chords are selected. The end result of the recipe will depend on the details of the chord selection just as an actual cake depends on the specific ingredients and tools used to implement the generic names and procedures used in the recipe.

\section{Releasing sticks: Probability 1/3 again}
\label{sec:release}

One sometimes witnesses doubts whether the various physical realizations of different chord selection method are equally valid.
Wang, for example, claims that only Jaynes' method is applicable because
\begin{quotation}
The fans of Bertrand's paradox assume carelessly that `random chords' that generated by any ``random'' method will be uniformly or homogeneously distributed in the circle. \cite[p.3035]{wang1}
\end{quotation}
This puzzling criticism is clarified in the later work of Wang and Jackson, who assert that
\begin{quotation}
[E]veryone agrees on what Bertrand-chords [chords generated randomly in a unit circle] are like - they are homogeneously or uniformly distributed over the circle. \cite[p. 76]{wang2}
\end{quotation}

But of course, not everyone agrees on this. Tissier, for example, also notes these facts but still admits that all procedures to select chords are equally valid, logically \cite{tissier}. It is, of course, precisely the point of Bertrand's argument – that the term ``random'' fails to specify the properties of the distribution sufficiently to determine it. Far from solving the paradox, to claim that ``random'' must mean some specific particular mathematical property of the distribution is merely strengthening Bertrand's point that different meanings of ``random'' will yield different distributions.

Wang and Jackson use the unspoken assumptions that people make, in their opinions. Thus, they say
\begin{quotation}
[W]hy do people think that this problem ought to have just one solution? The
reason can only be: Bertrand-chords are homogeneously distributed in people's
minds. \cite[p.77]{wang2}
\end{quotation}

The claim relies on what seems ``natural'', although neither I nor (I suspect, quite a few) several others think that this problem \textit{ought} to have just one solution, nor that it \textit{must} have uniformly distributed chords. Nevertheless, the question of a ``natural'' method of selecting chords at random does seem to arise here and there, even with those who recognize it is not sufficient grounds to justify any specific procedure. Thus Tissier considers that only solution B2 - the one ``proved'' by Jaynes - actually selects random chords directly \cite{tissier}. The others select some other properties that then determine the chords indirectly. Still, he recognizes that this does not invalidate the other procedures. Di Porto et al. refine Jaynes' throwing procedure to take into account the size of the straws and conclude ``we believe that our approach provides the natural solution to Bertrand's paradox'', although the only justification for this seems to be, if I read them correctly, that it is based on a physical experiment rather than randomly \textit{drawing} the chords  \cite{diporto1}. Jaynes himself seems to hold a similar position when he writes
\begin{quotation}
It will be helpful to think of this in a more concrete way; \textit{presumably}, we do no violence to the problem (i.e., it is still just as ``random'') if we suppose that we are tossing straws onto the circle, without specifying how they are tossed. \cite[p. 477, emphasis added]{jaynes}
\end{quotation}
But of course, as already pointed out by Shackel \cite{shackel}, Jaynes \textit{is} doing violence to the problem by choosing a specific physical procedure to implement it. Yet most subsequent authors (even those who ultimately reject Jaynes' attitude) accept that throwing straws is \textit{the} physical implementation of the problem, and that the principle of transformation groups thus ultimately does select a unique \textit{physical} solution to the problem (some reject the idea that this represents a solution to Bertrand's general query, however, e.g., \cite{shackel}).

Presumably, the reason why Jaynes' procedure seems to be so readily accepted as \textit{the} physical implementation has something to do with ``naturalness'', i.e., other selection methods seeming less directly a choice of actual chords. Selecting points on the circumference or throwing darts to specify a midpoint do not \textit{physically} generate a chord, although they do \textit{logically} select a unique entity. After the physical selection, the chord must be drawn through the selected points, or in the case of our impaled darts, the straw must be purposefully adjusted so that the selected point is indeed the midpoint. There is some artifice in these methods that may leave one doubtful as to their validity, though logically we recognize that they are acceptable, because they seem not to involve actual chords but something more abstract.

Jaynes' procedure is not without faults of its own, particularly because it is strictly valid only for infinite straws thrown form infinitely far away \cite{marinoff,diporto1,diporto2}. Additionally, one must take care to tweak them before the throw, lest there be a preferred orientation along the line of the arm holding the straw. Nonetheless, that Jaynes' calculation is supported by an experiment apparently lends it significant weight. The procedures described above can certainly be physically implemented, but they still suffer from not being direct selections of lines rather than points. In view of this possibly lingering doubt, I will now present a different implementation of method RB1, in which the chords are directly selected and which moreover only involves finite straws so that it is strictly practical, not merely as a limiting process.

Instead of a straw, I use a pointed stick (it must be rigid), of length $2R$ or slightly more. First, a random point is selected on the perimeter of a circle drawn on the floor (e.g., by using the stick as the needle of a spinner - this stage is not important because of the rotational symmetry of the problem). Let the angular position of its radius vector be $\psi$. At the chosen point, one then balances the pointed stick perpendicularly and releases it from rest. About half the times, the stick will fall outside the circle, counting as a failure (Jaynes' procedure also has similar failures), but when it falls across the circle, it thereby selects a chord. The second random variable, $\theta$, is the angle at which the stick falls, with respect, e.g., to the diameter passing through the point of release. We seek the probability distribution function $f_2\left(\psi ,  \theta\right)$.

This procedure is clearly another way to implement RB1. The reason why I describe it here is twofold. First, it lends itself very simply to experimental verification. More importantly, however, is that when one applies the principle of transformation groups, one discovers that the symmetries function quite differently from the case described in section \ref{sec:probthird}, even though they both represent implementations of the same solution. I can think of no better way to show that the application of symmetry conditions crucially depends on the details of the selection procedure.

We are considering the same three symmetries: rotation, scaling and translation. By the exact same argument as before, rotational symmetry implies that the PDF is independent of $\psi$. Note that the angle $\theta$ is unaffected by rotations since it is defined with respect to a specific diameter that rotates together with the circle. Thus, rotational symmetry does give us information, but not enough to fix the PDF, unlike in the case of spinners.

Scaling invariance, on the other hand, is again not expected to give us information, since angles are scale independent. It is instructive, however, to understand how scale invariance is to be applied in this case, because it nicely exemplifies how the ``same'' symmetry takes on different aspects according to the selection procedure that underlies it. Unlike the first two cases we saw, scaling invariance here cannot rely on comparing two concentric circles. The selection procedure assumes that we drop the stick from a point on the circumference, and concentric circles cannot touch. Instead, if we are to consider circles of different radii, they must be tangent to each other at the point of release of the stick. The situation is described in Fig. 3. It clearly shows, as expected, that every stick that generates a chord in one circle also generates a chord in the other, and that these two chords are either both longer than the sides of the respective inscribed triangles or both shorter. The angle $\theta$ that describes the orientation of the chord must be the same for both circles, and thus the PDF is trivially invariant whatever its form, as we knew beforehand. Nevertheless, it is worth noting, as mentioned, that the situation to which this symmetry applies - two tangent circles - is completely different to that to which it applied in Jaynes' procedure - two concentric circles. Thus, although the symmetry is ``the same'', meaning it is still scaling invariance, the specific form it takes depends on the chord selection procedure.

\begin{figure}
\label{fig:dropscale}
\includegraphics[scale=0.14]{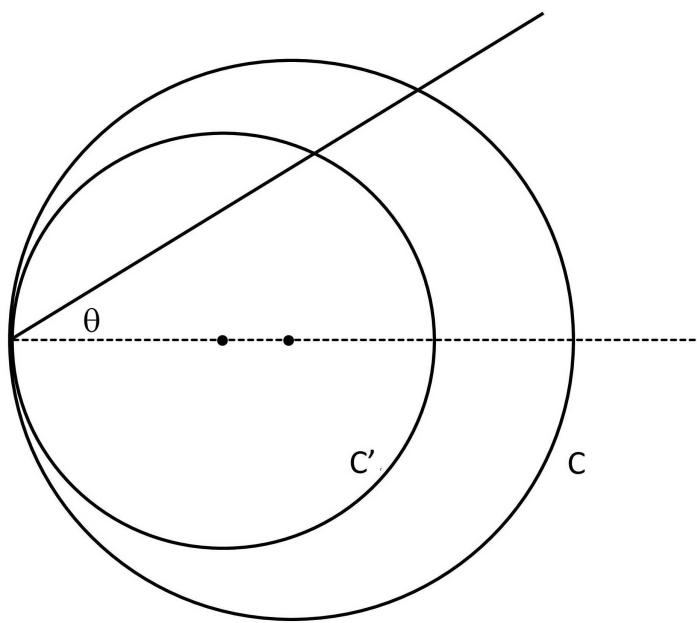}
\caption{Scaling invariance in the stick release procedure. The point of release is defined to be on the perimeter of the circle. A change of scale must imply that the smaller and larger circle are tangent at that point}
\end{figure}

The same applies to translation, which determines the PDF here. If a single stick is to create a chord in two circles, they must be tangent. Thus, not all translations are possible. Only those that leave one point on the perimeter of the circle in its place can be considered. The situation is described in Fig. 4, which shows two circles whose centers are a distance $b$ from each other, so that they share the point of origin of the chord. The diameters of the two circles that pass through that point form an angle $\phi$, and the angle of the chord in the translated circle, $\theta'$, is given by
\begin{equation}
\theta' = \theta - \phi
\end{equation}

\begin{figure}
\label{fig:droptrans}
\includegraphics[scale=0.14]{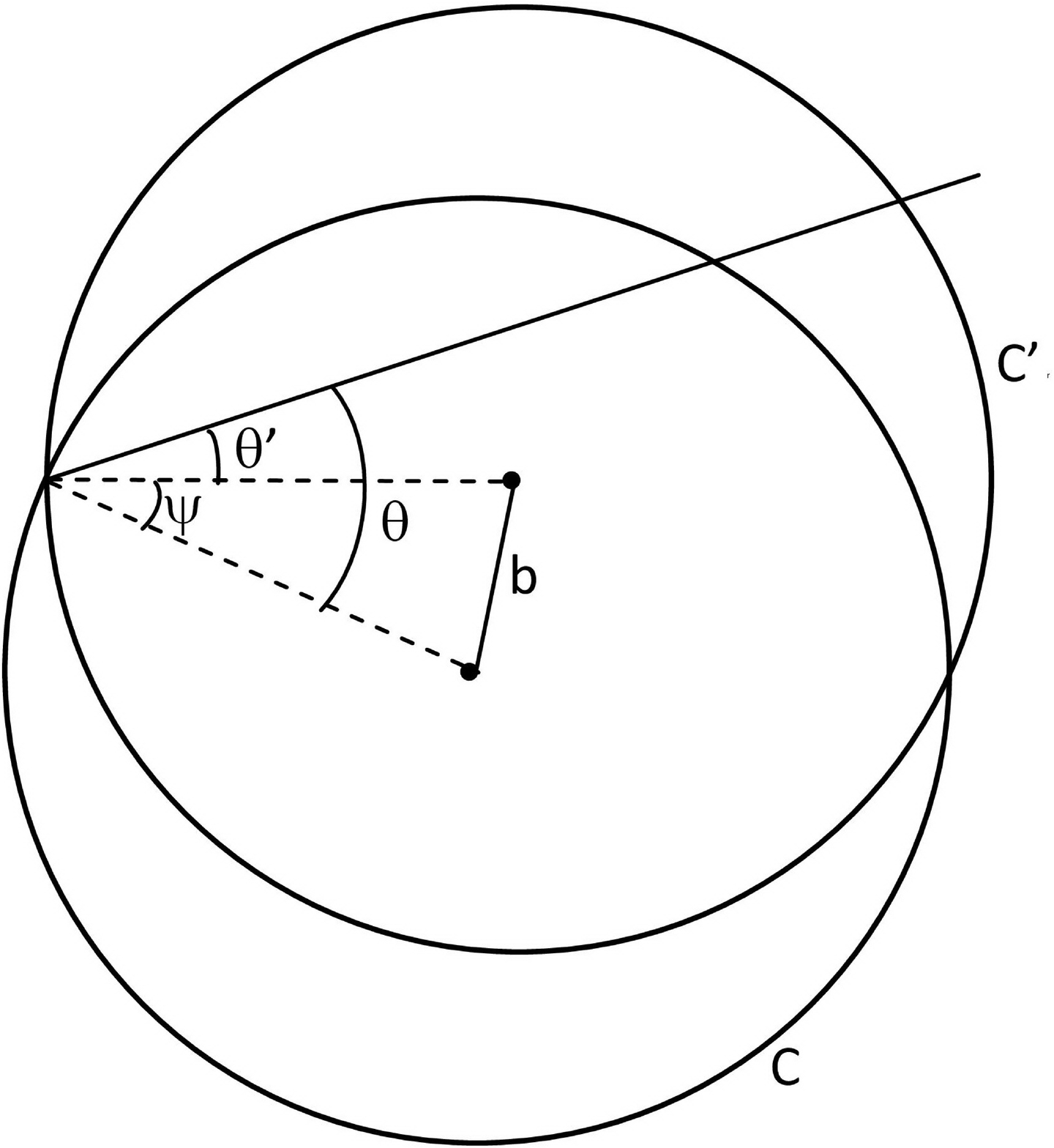}
\caption{Translational invariance in the stick release procedure. One may translate the circle provided the point of release remains on the perimeter. This is equivalent to a rotation around that point by an angle $\psi$. The stick then defines two chords in the two translated circles, with different orientations}
\end{figure}

Since $d\theta' = d\theta$, invariance of the probabilities to such translations must mean now that
\begin{equation}
f_2(\theta) = f_2(\theta') = f_2(\theta - \phi)
\end{equation}
Since this must hold for any $\phi$, $f_2$ must be the constant
\begin{equation}
f_2(\theta, \psi) = \dfrac{1}{2 \pi^2}
\end{equation}

The Bertrand probability is given by a calculation almost identical to that of Eq.(\ref{eq:spinprob}) and is similarly 1/3. Once again, this is the result of applying the principle of transformations groups. It differs in its underlying realization from the others cases we have considered, although it gives the same numerical answer as the spinner solution. It represents a different physical implementation, however, and the form of the symmetry relations it entails are consequently also different.

To verify the result, I drew a circle on a piece of paper, which I then dropped from a certain height, giving it a twist in the process. When the paper landed on the floor, I placed a pencil inside it and twirled it. The point on the circumference at which the pencil pointed after coming to rest was chosen as one end of the chord. The aim of this procedure was to randomize as much as possible both the position of the piece of paper and my own, in order to counteract irregularities in the floor as well as any preferences my muscles may have. I then stood a thin pointed stick on its end at the selected point and released it from as perpendicular a position I could manage. Because the procedure is obviously very sensitive to slight tilts in the positioning of the stick, and thus to any systematic errors in setting it upright, I repeated the experiment more times than Jaynes performed his.

Out of 700 attempts, 363 were successes (i.e., the stick fell over the circle), a 0.518 probability. I used this number to estimate whether I had a significant systematic error in positioning the stick. Although one might have expected to be closer to the 0.5 expected figure, I consider the result to be fairly satisfactory. Out of the successful releases, the chord was longer than the side of the inscribed triangle 123 times, or a proportion of 0.339, quite close to the expected theoretical value 1/3.

Although more sensitive to systematic errors that Jaynes', this procedure likewise directly selects chords (rather than some abstract characteristic like the midpoint) and is therefore equally ``natural''. We have thus another experimental method of selecting chords, which yields a different result from Jaynes'. This result confirms a theoretical calculation based on the very same invariance properties Jaynes used, properly understood in the context of the procedure, viz., rotational, scaling and translational symmetries. Thus, we see again that the principle of transformation groups can generate a different solution, and that this solution is equally verified experimentally. More methods of directly generating chords are available and can be similarly empirically verified. They yield still different values from the ones considered here and represent additional solutions beyond the original three of Bertrand \cite{drory}.

\section{Discussion}
\label{sec:conclusions}

In his discussion of the principle of indifference as a heuristic tool in physics, Gillies imagines the following scenario:

\begin{quotation}
[C]onsider[...] again Jaynes's analysis of the random chord case and the confirming experiment he performed. The same conclusion could have been reached by another scientist (Mr K say) following a different route. Let us suppose that Mr K applies the Principle of  Indifference to the random chord case, but initially only the third approach [B3] (which yields P(CLSE) = 1/4) occurs to his mind. He works out the full distribution of chord lengths on this approach and then tests the distribution using exactly the same experiment as Jaynes. In the case of Mr K, however, the experiment disproves his hypothetical distribution. In the face of this refutation, Mr K analyses the problem further. He hits on the other two ways of applying the Principle of Indifference, and he also thinks of the invariance requirements which suggest that the first approach is the best of the three. In this way he explains the result of his experiment successfully.
\end{quotation}

This is certainly a possible sequence of events, but it is not the only one, and although I agree with Gillies' last sentence, I have reservations about the rest. As I read Gillies' story, he imagines that Mr. K uses only the principle of indifference in his initial calculations, but not symmetry arguments. He can then derive the distribution of chord lengths from the assumption that the midpoint of the chords is selected from a uniform distribution over the area of the circle. So far so good, but why would Mr. K then use Jaynes' experiment to test his distribution? To be sure, it may be that this is the only option that comes to his mind, but as we have seen, it is not the only one available. Contrary to an apparently received opinion, there are other methods of experimentally selecting chords at random, and these yield different Bertrand probabilities (I do mean selecting \textit{chords}, rather than some characteristics that can be used to later draw the chord). This means that when Mr K obtains a discrepancy and analyzes the problem further, he may not necessarily hit on the two other ways to apply the principle. Instead, he may hit on another experimental method to test his distribution, and it may very well be that it will confirm his calculation.

The most important point, however, is that supposing Mr K hits on the other Bertrand solutions as well as on the idea of using invariance requirements, the result will not be what Gillies thinks. In fact, each of these solutions can be supported by invariance requirements and even by the very same requirements, in the sense that they will be all called rotation, scaling and translation invariance. But the mathematical restrictions they impose on the PDF's differ because just like the principle of indifference, they can be applied in different ways, depending on how one selects the random chords in the circle.

If the hypothetical Mr K is a scientist, as Gillies describes him, rather than a philosopher, he will likely still be satisfied, for he will still be able to explain the results of his experiments. He will realize that he has several available empirical procedures, that each implies certain mathematical requirements that follow from symmetry considerations, and that for each procedure, he can actually calculate theoretically the PDF and find that it describes the results of the experiments. That the calculations all yield different values will not surprise him, since he will expect that different procedures (with different implementations of the symmetries) should yield different results, both theoretically and experimentally.

If Mr K is a philosopher, on the other hand, he may remain unsatisfied, but he has made some progress nevertheless. True, he still has a choice among various positions. He can take Bertrand's problem to be ill-posed but resolvable into several well-posed problems once the selection procedure is set, following Marinoff \cite{marinoff}, or he can insist that there still must be a sense in which ``random'' means something specific, even if only on the meta-level, like Aerts and Sassoli de Bianchi \cite{aerts}, or he can also consider that the problem remains an unsolved challenge, as does Shackel \cite{shackel}. What he has gained, however, is the definite understanding that the principle of transformation groups does not make the problem well-posed, and that well-posing strategies that rely on such symmetry considerations ought therefore to be rejected. Whatever he believes of the principle of indifference should equally hold for the principle of transformation groups. If he believes that the principle of indifference is only applicable after the problem has been separated into well-posed alternatives, then so is the principle of transformation groups. If he believes that the principle of indifference fails, then so does the principle of transformation groups.

This is not to say that Jaynes' principle doesn't have its uses. It will not let us obtain a definite answer where the principle of indifference leaves us confused, but it can be a powerful heuristic and formal device to guide us to the correct PDF once we have decided what the proper selection procedure is. Consider Jaynes' own procedure of throwing straws, and let us imagine that the logical order of the questions is reversed. In other words, let us assume that the question we ask is ``\textit{given} this selection procedure, what is the correct PDF for the chord's length?''. The principle of indifference is hard to apply here, because the procedure does \textit{not} directly select the midpoints of the chord, although these are the parameters we use. Instead, it determines these coordinates indirectly from the position of the straw. Thus, the random variables we use are not directly selected and the principle of indifference should not be applicable to them, as indeed it isn't. 

When facing such a problem, one can try two approaches. The first is to transform the problem into something to which the principle of indifference is directly applicable. This suggests that the order of Jaynes' analysis is backwards. One starts with a specific experimental procedure, then seeks a more abstract formulation. In this case, the argument of Tissier, already alluded to above several times, would suggest that the problem is isomorphic to that of randomly selecting a chord from a set of  parallel lines \cite{tissier}. This in turn would bring us to formulate Bertrand's problem and the solution RB2 (or B2). To this, the principle of indifference suggests a solution and a Bertrand probability of 1/2.

The other approach would probably be considered more ``physical''. It is here that the principle of transformation groups shows its utility. Instead of seeking the correct sample space on which to assume a uniform probability distribution, we impose physical conditions on the problem in the form of symmetry requirements. These are not abstract properties, however, but specific mathematical conditions derivable from the procedure we are considering. For that particular procedure, the principle of transformation groups would then yield the PDF on a basis that would not depend on analogies with more abstract formulations. It is precisely the close link of these symmetries to the details of the procedure that gives the argument strength. It is \textit{because} it depends on the physical details of the selection that we can trust the derivation. Analogies and abstractions are vulnerable to the adequacy argument. We can never be sure that in transforming our problem into something supposedly equivalent we have not inadvertently changed some crucial aspect. For those problems where the application of the principle of indifference is unclear or the correct sample space not immediately apparent, the principle of transformation groups offers an alternative grounded in the physical properties of the selection procedure. It it to these practical problems that it should prove useful, rather than to general philosophical discussions.

\end{document}